\documentstyle{mn2e}
\def\etal{{\it et al.~}}

\def\half  	{\textstyle {1 \over 2} \displaystyle}
\def\third  	{\textstyle {1 \over 3} \displaystyle}
\def\4ov3  	{\textstyle {4 \over 3} \displaystyle}
\def\8ov11  	{\textstyle {8 \over 11} \displaystyle}
\def\15ov16  	{\textstyle {15 \over 16} \displaystyle}

\def\beq{\begin{equation}} \def\eeq{\end{equation}}
\def\bea{\begin{eqnarray}} \def\eea{\end{eqnarray}}

\def\AM{ 	\hat{\rm A} } 	
\def\qB{ 	\hat{q}_{_{\hat{\bf B}}} } 	
\def\hchi{ 	\hat{\chi} }

\def\sD{{\cal D}}

\def\cH{{\cal H}}
\def\cO{{\cal O}} 

\def\ccH{ c_{_\cH}}
\def\Valf{{\it v}_{_{\rm Alf}}}
\def\Alf{{\rm Alfv\'en}~}
\def\Oor{{\rm Oort}}
\def\AG{{\rm Araya-G\'ochez\,}}

\def\Ptot{p_{\rm tot}}
\def\Prg{p_{\rm r + g}}
\def\Prad{p_{\rm rad}}
\def\Pgas{p_{\rm gas}}

\def\PB{p_{_{\rm \bf B}}}

\def\bx{ {\bf x}} 	\def\bk{ {\bf k}}	\def\bn{ {\bf n}}
\def\bB{ {\bf B}} 	 	\def\bH{ {\bf H}}
\def\bv{{\bf v}} 	\def\bV{ {\bf V}} 	\def\bJ{ {\bf J}} 	
\def\bOm { {\mbox{\boldmath $\Omega$}} }
\def\bxi { {\mbox{\boldmath $\xi$}} }
\def\unvc{ {\mbox{\boldmath $1$}} }
\def\bbeta{ {\mbox{\boldmath $\beta$}} }

\def\Del { {\mbox{\boldmath $\nabla$}} }
\def\del { {\partial}}
\def\tdel{\tilde{\partial}} 	
\def\tDel{\tilde{\Del}}

\def\hom{\hat{\omega}}	 
\def\tom{\tilde{\omega}} \def\tk{\tilde{k}} 	\def\teta{\tilde{\eta}}	
\def\AOM{|\omega_{_{\rm MRI}}|} 	
\def\tAOM{|\tilde{\omega}_{_{\rm MRI}}|}
\def\hAOM{|\hat{\omega}_{_{\rm MRI}}|}

\def\mfp{\ell_{\rm mfp}}

\def \ltaprx {\lower .1ex\hbox{\rlap{\raise .6ex\hbox{\hskip .3ex
	{\ifmmode{\scriptscriptstyle <}\else 
		{$\scriptscriptstyle <$}\fi}}}
	\kern -.4ex{\ifmmode{\scriptscriptstyle \sim}\else 
		{$\scriptscriptstyle\sim$}\fi}}}
\def\gtaprx {\lower .1ex\hbox{\rlap{\raise .6ex\hbox{\hskip .3ex
	{\ifmmode{\scriptscriptstyle >}\else 
		{$\scriptscriptstyle >$}\fi}}}
	\kern -.4ex{\ifmmode{\scriptscriptstyle \sim}\else 
		{$\scriptscriptstyle\sim$}\fi}}}

\begin{document}

\title{Radiative Heat Conduction and the Magnetorotational Instability} 

\author [R.A. Araya-G\'ochez and E.T. Vishniac]
	{Rafael A. Araya-G\'ochez$^1$
	and Ethan T. Vishniac$^2$ \\
$^{1}$Theoretical Astrophysics, 
California Institute of Technology, MC 130-33, Pasadena, CA 91125 
\, {\bf arayag@tapir.caltech.edu} \\
$^{2}$Department of Physics and Astronomy, The Johns Hopkins University, 
Baltimore, MD 21218
\,\,  {\bf ethan@pha.jhu.edu} } 

\onecolumn
\date{}
\pubyear{2003}
\maketitle

%##################################L100###################################

\begin{abstract} 
 A photon or neutrino gas--semi-contained by a baryonic species 
 through scattering--comprises a rather peculiar MHD fluid where 
 the magnetic field is truly frozen only to the co-moving volume 
 associated with the mass density.  
 Although radiative diffusion precludes an adiabatic treatment of
 compressive perturbations,
 we show that the energy equation may be cast in ``quasi-adiabatic" form
 for exponentially growing, non-propagating wave modes.
 Defining a generalized quasi-adiabatic index 
 leads to a relatively straightforward dispersion relation for 
 non-axisymmetric magnetorotational modes in the horizontal regime
 when an accretion the disk has comparable stress contributions
 from diffusive and non-diffusive particle species.
 This analysis is generally applicable to optically thick, 
 neutrino-cooled disks
 since the pressure contributions from photons, pairs and neutrinos, 
 all have the same temperature dependence 
 whereas only the neutrino component has radiative heat conduction 
 properties on the time and length scales of the instability.
 We discuss the energy deposition process 
 and the temporal and spatial properties of the ensuing turbulent disk
 structure on the basis of the derived dispersion relation. 
\end{abstract}    
%#########################################################################
%#########################################################################

\section{Preliminaries}
\label{sec:Prelim}

 	Understanding the magnetorotational instability
 is key to the development of realistic models of accretion.
 As a physical process, the MRI justifies two 
 {\it sine qua non} ingredients of accretion disk theory:
 entropy generation from the differential shear flow 
 and turbulent angular momentum transport. 
 Consequently, clarifying the dynamics that leads to MRI initiated 
 turbulence in a few relevant astrophysical regimes is paramount 
 in uncovering how the accretion process takes place.

 	In the context of accretion onto compact objects,  
 the stress associated  with a radiative particle species 
 often dominates the dynamics at high temperatures but 
 our understanding of its effect on MHD processes is incomplete at best.
 Agol \& Krolik (1998) pointed out that compressive radiation-MHD wave 
 modes will be strongly suppressed by diffusive loss of pressure 
 support in a range of wave numbers:
 ${ c_{\rm ph} / c} < \tk^{-1} < 1$,
 where $\tk \equiv \mfp \, k$ 
 represents the wave number normalized to the mean-free-path and 
 $c_{\rm ph}$  
 is the phase speed of compressive waves in the single fluid view.
%-- (with $c_r \equiv \4ov3 \Prad/\rho$,
%-- $c^2_{\rm ph} = 3\{3c_r^2 + c^2_g + \Valf^2 \}$
%-- at the threshold between diffusive and optically thin regime; and
%-- where $c_g$ and $\Valf$ are the sound speed of the material gas
%-- and the \Alf speed).
 Such analysis is applicable to accretion disks at large
 wave numbers where the effect of inertial forces 
 and stratification are minor.  
 For the MRI, however,
 both inertial and stratification effects play key roles
 in dictating the dynamics and polarization of the modes.

 	Blaes \& Socrates (2001) have reported on a radiation-MHD 
 dispersion relation for axisymmetric modes that comprises 
 the MRI under rather general circumstances.
 Our analysis is different from theirs mainly in the consideration 
 of a purely toroidal field where the compressibility of the 
 non-axisymmetric modes leads to the largest impact 
 from radiative diffusive damping.
 Furthermore, 
 aiming to understand ultra-relativistic flows very near black holes,
 we emphasize toroidal mode dynamics because these modes may become 
 predominant when shear effects begin to overwhelm other inertial 
 accelerations leading to fastest growing modes at large azimuthal 
 scales (\AG 2002).  
 The scale and degree of intermittency in the turbulent flow bear 
 a direct impact on gravitational wave emission models from 
 hyper-accreting black holes (\AG 2003).  
 For a relativistically hot accretion disk
 with comparable vertical pressure support 
 from diffusive and non-diffusive species
 (see \S \ref{sec:Scales}),
 we solve for an approximate dispersion relation involving 
 Lagrangian displacements in the {\em horizontal regime}
 ($\xi_\theta/\xi_r \simeq k_r/k_\theta \rightarrow \emptyset$).
 This simplification affords a lowering 
 of the rank in the dispersion relation to second order 
 (in $\omega^2$) which, in turn, 
 enables a relatively straightforward approximate treatment of the 
 effects of radiative heat conduction
 (in contrast with Blaes and Socrates' more complex but exact, 
 relation for axisymmetric modes).
% The Lagrangian displacement formalism also 
% makes manifest a very useful connection between the compressibility  
% and the polarization of the fastest growing MRI modes (Foglizzo 1995).

%#########################################################################
\section{Radiative Viscosity {\it vs} Diffusion} 
 \label{sec:Scales}

 	The half optical depth to scattering of a 
 standard radiation-pressure--dominated $\alpha$-disk 
 is related to the vertical scale height, $\cH$, 
 and to the angular rotation profile, $\Omega(r)$, through
\beq
 \tau_{\rm disk} =  - \frac {c} {2 \alpha \AM \Omega \cH}
\label{eq:DiskDepth} \eeq
 where \Oor's A parameter is normalized 
 $\AM \equiv \half {\rm d}_{\ln r} {\ln \Omega}$
 ($\doteq -3/4$ for a non-relativistic, Keplerian flow).
 This relation
 ensues from balancing the local heating rate,
 $\propto \dot{M} \Omega^2 \AM$,
 with the local cooling rate from radiative diffusion,
 $\propto \nabla_\tau p$; 
 while eliminating the accretion rate in favor of the vertically 
 integrated component of the viscous stress responsible for angular momentum
 transport $\propto \alpha \, \cH p$.
 Eq \ref{eq:DiskDepth} is thus 
 explicitly sensitive only to the local nature of the 
 cooling  
 (although it is implicitly subject to a suitable vertical 
 gradient of heat deposition; see, e.g., Krolik 1999).
	
 	In the neutron-rich, neutrino cooled gas associated with a 
 hyper-accreting black hole (see, e.g. Popham \etal 1999), 
 the pressure contributions from radiation, pairs and neutrinos, 
 all have the same temperature dependence, $\propto a/6~T^4$,
 with relative contributions 
 varying only by internal degrees of freedom 
 times particle statistics factors:   
 2 $\times $ 1, 4 $\times$ 7/8, and 6 $\times$ 7/8, 
 respectively (with only one helicity state per neutrino).
 Thus, the neutrino pressure is never greater than $\simeq p_{\rm thermal}$
 where $p_{\rm thermal} = 11/12 a T^4$ includes the pressure
 from photons and pairs. 
 Yet, only the former species has radiative diffusion properties 
 on a dynamical time scale. 
 Thus, in this setting the neutrinos are the lone diffusive species
 and their stress is identified with $\Prad$ below 
 (e.g., radiative = diffusive species throughout this paper).

 	Partial pressure support by the non-diffusive components 
 (baryons, toroidal field and the thermal component)
 does not modify Eq \ref{eq:DiskDepth} 
 as long as the contribution to the disk's flare from this extra pressure
 is inversely proportional to the mass density decrease from the same, 
 e.g., 1D expansion.  Consequently, 
 for a neutrino cooled $\alpha$-disk, 
 a replacement in the source of opacity by neutrino scattering
 in the non-advective
 accretion regime:
 $0.1 \, {\rm M}_\odot \sec^{-1} \leq \dot{\rm M} 
	\leq 1 \, {\rm M}_\odot \sec^{-1}$
 (Di Matteo \etal 2002), 
 is all that is needed for Eq \ref{eq:DiskDepth} to hold. 

 	On the other hand, 
 the ``parallel" size the of fastest growing MRI eddies  
 is $l_{\rm eddy} = k_{\rm MRI}^{-1} \simeq \Valf / \Omega$.
%-- random walk is $c/\tau_{\rm eddy}$.
 For magnetic angular momentum transport such that  
 $\alpha \simeq (\Valf / \ccH)^2$, 
 the {\it isotropic} diffusion time through these eddies, 
 $t_{\rm diff} \sim \Omega^{-1}$, 
 turns out to be similar to the timescale for fastest MRI modes to develop
 $t_{\rm MRI} \simeq -{\rm A}^{-1}_\Oor$ (see \S \ref{sec:Impact}).
 Moreover, the optical depth through these eddies is 
 (with $\ccH^2 \equiv \Ptot/\rho$)
\beq
 	\tau_{\rm eddy} \sim {\Valf \over \ccH} \times \tau_{\rm disk}
			\sim - {1 \over {2 \AM \sqrt{\alpha}}} \, 
			{c \over \ccH} \gg 1. 
\eeq 

 	Since Compton/neutrino drag go as 
 ${\bf f}_{\rm drag} 	\propto 
			\tk^2 \, \delta \bbeta$
 and $\tk^2 \simeq \cO (\alpha \, c^2_{_{\cH}}/c^2)$,
 these estimates indicate that the effect of photon/neutrino viscosity  
 is negligible for the optically thick eddies of a
 radiatively cooled disk.  
%Only radiative diffusion has a significant impact on the MRI.

%######################################################################
\section{A Lagrangian formulation of the MRI in soft media}
 \label{sec:LagForm}

 We take a semi-local approach by computing instantaneous 
 (co-moving) growth rates in the limit $k_r/k_\theta \rightarrow 0$ 
 and referring to these as the fastest growing modes 
 on the implicit understanding of their transient nature
 (since 
 $k_r(t) = k_{0r} - [{\rm d}_{\ln r} \Omega ] \, k_\varphi t$).
 In terms of a local, co-moving observer, 
 azimuthal wave numbers are no longer discrete 
 and consequently, neither are the co-moving frequencies
 (Ogilvie \& Pringle 1996).  Further, 
 we discard field curvature and radial gradient terms.
 A simple meridional stratification profile 
 sets the physical scale-length of the problem: 
 d$_z \ln \rho = 1/\cH$, %-- ~(= - g_z/(c^2_s + \Valf^2))$,
 with gas, radiative and magnetic pressures
 tracking the unperturbed density profile.
% The precise form of stratification profile 
% is inconsequential when $k_r/k_z \rightarrow \emptyset$ (see below).

 	The first order relation between Lagrangian, $\Delta$, 
 and Eulerian, $\delta$, variations
 (see, e.g., Chandrasekhar \& Lebovitz 1964, 
 Lynden-Bell \& Ostriker 1967)
 is 
\beq
 {\Delta} = \delta + \bxi \cdot \Del.  
\label{eq:EulLagRel} \eeq 
 The Euler velocity perturbation, $\bv \equiv \delta \bV$, is thus 
 related to the Lagrangian displacement, $\bxi$, through
\beq
	\bv = \{\del_t + \bV \cdot \Del\} \, \bxi - (\bxi \cdot \Del) \bV
	\rightarrow \, i \sigma \bxi - \xi^r \Omega_{,r} \unvc_\varphi.
\label{eq:DynSwi} \eeq
%-- whilst denoting
%-- ${\Delta} \bV \equiv {\rm d}_t \bxi$.
 The algebraic relation follows from the assumption of 
 differential rotation and from writing 
 exp$\;i(\omega t + m \varphi + k_z z)$ 
 dependencies for $\bxi$, while 
 $\sigma \doteq \omega + m\Omega$ 
 denotes the co-moving frequency of the perturbations. 

	These geometrical equations have their more traditional equivalents 
 in the so-called shearing sheet approximation where a co-moving,
  ``locally Cartesian frame" 
 $(\hat{r}, \hat{\varphi}, \hat{\theta}) \rightarrow (x,y,z)$,
 is used along with the linearized shear velocity field, 
 $\bV(x) = [{\rm d}_{\ln r} \Omega] \, x \unvc_y$,
 to treat the problem locally while introducing the coriolis terms by hand.
 (The reader is invited to examine a companion paper: \AG (2002), 
 where the inertial terms are induced in a fully covariant, geometric way).

 	The linearized equations of motion for the displacement vector
 correspond to the Hill equations (Chandrasekhar 1961, Balbus \& Hawley 1992)
\beq
 	\ddot{\xi}_r - 2\Omega \, \dot{\xi}_\varphi = 
		-4 {\rm A}_\Oor \Omega \, \xi_r  + {f_r \over \rho},
~~~~	\ddot{\xi}_\varphi + 2\Omega \, \dot{\xi}_r = {f_y \over \rho}, 
~~~{\rm and}~~	\ddot{\xi}_\theta = {f_\theta \over \rho},
\label{eq:Hill} \eeq
%\begin{eqnarray}
% 	\ddot{\xi}_r - 2\Omega \, \dot{\xi}_\varphi &=& 
%		-4 {\rm A}_\Oor \Omega \, \xi_r  + {f_r \over \rho},	\cr
%&&\cr
% 	\ddot{\xi}_\varphi + 2\Omega \, \dot{\xi}_r &=& {f_y \over \rho}, 
%~~~{\rm and}~~	\ddot{\xi}_\theta ~~=~~ {f_\theta \over \rho},
%\label{eq:Hill} \end{eqnarray}
 where each over-dot stands for a Lagrangian time derivative 
 and {\bf f}, for the sum of body forces.

 	The Lagrangian perturbation of mass density 
 and the Eulerian perturbation of the field 
 follow straightforwardly from Eq [\ref{eq:EulLagRel}], 
 and from mass and magnetic flux conservations 
 (under ideal MHD conditions,
 notwithstanding the peculiar nature of the fluid under study)
\beq
	{\Delta \rho \over \rho} = - \Del \cdot \bxi, ~~~~~~
	\delta \, \bB =  \Del \times ( \bxi \times \bB).
\label{eq:MasterEqs} \eeq
 Using the latter relation,	
 the Lorentz force in the co-moving frame is readily laid out 
\bea
	\delta {1 \over {4\pi\rho}} \bJ \times \bB &=& \Valf^2
	 \,\, \times \,\,
 	\left\{	\,\,  \Del (\Del \cdot \bxi) 	
		+ \nabla^2_{\hat{\bB}} \bxi \right.	
%\cr &&	
	\left.	- \nabla_{\hat{\bB}} \Del (\unvc_{\hat{\bB}} \cdot \bxi)
		- \unvc_{\hat{\bB}} \nabla_{\hat{\bB}} (\Del \cdot \bxi) 
		\right\} \cr
&&\cr
&& \!\!\!\!\!\!\!\!\!\!\!\!\!\!
	 \stackrel{\nabla \rightarrow i\bk}{\longrightarrow} 
	-\Valf^2 \times
	\{	(k_i \xi_i - k_{\hat{\bB}} \xi_{\hat{\bB}}) \bk
	+	 k^2_{\hat{\bB}} \bxi 	
	- 	\unvc_{\hat{\bB}} k_{\hat{\bB}} k_i \xi_i.
	\} 	
\label{eq:Loren_F} \eea 
%-- NOT a Fourier transform of \bxi? 
 where $\unvc_{\hat{\bB}}$ is a unit vector in the direction of 
 the unperturbed field and where the scalar operator 
 $\nabla_{\hat{\bB}} \equiv  \unvc_{\hat{\bB}} \cdot \Del$.
 Note that the term 
 $(k_i \xi_i - k_{\hat{\bB}} \xi_{\hat{\bB}}) \doteq k_\perp \xi_\perp$
 may be interpreted as a restoring force due to the 
 compression of field lines (Foglizzo \& Tagger 1995).
  
	Next, we look in detail at the non-magnetic stress terms.

 	In the single fluid view, the Lagrangian variation 
 of the specific pressure gradient, $\rho^{-1} \Del \Prg$,
 contains two terms (see, e.g., Lynden-Bell \& Ostriker 1967):
%-- Lynden-Bell. D. \& Ostriker, J. 1967, MNRAS 136, 293
 one $\propto \Delta \rho^{-1}$,
 and another one $\propto \Delta \Del \Prg$.
 In terms of the displacement vector, the first term is  
 proportional to the equilibrium value of $\Del \Prg$ 
 which is negligible in the local treatment 
 ($\propto$ radial gradient when $\xi_\theta/\xi_r \rightarrow \emptyset$).
 For the same reason, the Eulerian and Lagrangian variations 
 of the specific pressure gradient are nearly identical.
 
  	The ``thermodynamic" pressure term is thus given by
\beq
 	- \delta \left( {1 \over \rho} \Del \Prg \right) 
	= \Gamma \, {\Prg \over \rho} \Del (\Del \cdot \bxi) 
	\, \stackrel{\nabla \rightarrow i\bk}{\longrightarrow} \,
	c_s^2 \, \bk \, k_i \xi_i,
\label{eq:Therm_F} \eeq
 where, for a heterogeneous fluid, 
 $\Gamma (\equiv d_{[\ln \rho]} \ln p)$  
 represents a generalized adiabatic index 
 (see, e.g., Chandrasekhar 1939, Mihalas \& Mihalas 1984).
%In \S \ref{sec:DifAdiaInd}, we show that
%when the lone dynamical role of the material gas is to contain the 
%radiative component through scattering, e.g. for $\Pgas \ll \Prad$, 
%the quasi-adiabatic index comprises a real function of the perturbation 
%scales; c.f. Eq [\ref{eq:QuasiAdiab}].
%On the other hand, 
%when $\Pgas \simeq \Prad$, this index is
%generally complex; c.f. Eq [\ref{eq:DmpAdi}].

	Assembling {\bf f} from 
 Eqs [\ref{eq:Loren_F} \& \ref{eq:Therm_F}],
 and plugging this form into Eqs [\ref{eq:Hill}] yields
\beq 						
	\ddot{\bxi} + 2 \, \bOm \times \dot{\bxi} =
%-- &=& 
	\{ 	c_s^2 			\, \bk 
	+ \Valf^2 ( \bk - k_{\hat{\bB}}	\, \unvc_{\hat{\bB}}) \} 
	(\bk \cdot \bxi) 
%-- \cr &&\cr &+&	
	+ \Valf^2 (	k^2_{\hat{\bB}} \, \bxi
	- k_{\hat{\bB}} \xi_{\hat{\bB}} \, \bk)
	- 4{\rm A} \Omega \, \xi_r 	\, \unvc_r.
\label{eq:k-EOM} \eeq
 This equation agrees with the matrix de-composition of 
 Foglizzo \& Tagger (1995) for $\Gamma = 1$.
 Moreover, these authors also pointed out that  
 because of vertical stratification and rotation,
 the polarization of the slow MHD (e.g., MRI) modes  
 obey 
\beq 	
 {\xi_\theta \over \xi_r} \simeq \cO \left( {k_r \over k_\theta} \right)
\label{eq:Polariz} \eeq
 when $k_r/k_\theta \rightarrow \emptyset$.
 Eqs [\ref{eq:k-EOM} \& \ref{eq:Polariz}] impose an anisotropy
 constraint on the components of the Lagrangian displacement 
\beq
 k_\perp \xi_\perp = - {\Gamma \over {\Gamma + 2 \Theta}} 
			\, \,	k_\parallel \xi_\parallel 
		~~ \equiv - \Lambda \, k_\parallel \xi_\parallel,
\label{eq:Lamb} \eeq
 where $\Theta \equiv \PB/\Prg$. %-- = {\Gamma \over 2} \Valf^2/c^2_s$.  

 	In this horizontal regime, with $k_z \xi_z$ finite,
 the dispersion relation derived from 
 Eq [\ref{eq:k-EOM}] and from the above constraint reads 
\beq 		\hom^4 
	- \{ (	\Lambda + 1) \qB^2 + \hchi^2 \} \, \hom^2
	+ 	\Lambda \, \qB^2 \{  \qB^2 +  4 \AM \}
	= \emptyset,
\label{eq:disper} \eeq
 where all frequencies  
 are normalized to the local rotation rate, 
 $\hchi^2 \equiv 4(1+\AM)$ is the squared of the epicyclic frequency, and 
 $\qB \equiv ({\bk} \cdot {\bf v}_{\rm Alf}) / \Omega$ is a 
 frequency related to the component of
 the wave vector along the field (in velocity units). 
 This expression matches the form given by 
 T. Foglizzo (1995).

	In terms of the function, $\Lambda (\Gamma,\Theta)$,  
 we find the fastest growing wave numbers to be given by
\beq
	\qB^2 =	- 2 \AM 
	+ ( {\textstyle {{1 + \Lambda}\over{2\Lambda}} \displaystyle} )
	\times	  \left\{ -{2 \Lambda \AM^2 \over \sD} \right\},
 ~~~{\rm where}~~~ 
	\sD \equiv 1 
		+  ({\textstyle { {1 - \Lambda}\over2 }\displaystyle}) \AM
		+  \sqrt{1 + (1 - \Lambda) \AM},	
\label{eq:qB-om} \eeq
 and 
 where the expression in the curly brackets corresponds to the growth 
 rate of the modes. 

 	Notably, 
 the compressibility of non-axisymmetric, MRI modes is imprint on 
 deviations of $\Lambda$ from unity 
\beq
	\frac{\Delta \rho} {\rho} = 
	(1-\Lambda) (i k_\parallel) ~ \xi_\parallel,
\label{eq:Compress} \eeq
 e.g., the degree of compression gets stronger with field 
 strength and, naturally, with a softer equation of state.

%##############################################################################
\section{A Quasi-Adiabatic Index for Wave Phenomena}
 \label{sec:DifAdiaInd}

	In a heterogeneous fluid composed of a neutral baryonic plasma
 (with ideal gas properties) 
 plus a neutrino and/or photon component, ideal MHD dictates that the 
 field lines remain frozen to the charged massive component 
 while pressure perturbations associated with such material gas 
 must track density perturbations on all scales. 
 On the other hand,
 collisional coupling between the gas and the diffusive species 
 translates into a scale dependent containment of the pressure perturbation
 associated with the radiative species.
 Consequently, 
 acoustic wave modes partially supported by radiative stress
 will be damped from a non-adiabatic loss of pressure support 
 when the diffusive and wave time scales are comparable 
 (Agol \& Krolik 1998). 
 Such phenomenology can be quantitatively implemented by considering the 
 mathematical equivalent to the equation of state for the 
 diffusive component of the fluid. 

	 We start with the co-moving frame, frequency-integrated,
 radiative transfer equation--corrected to first order in the 
 fluid's motion, $\bbeta$, and 
 with isotropic, elastic scattering as the lone source of opacity:
\beq
 	\tdel_t I(\bn) + \bn \cdot \tDel I(\bn) =
	(1 + 3 \bbeta \cdot \bn) J - 2 \bbeta \cdot \bH
	- (1 - \bn \cdot \bbeta) I(\bn).
\label{eq:Transf} \eeq
 A tilde means normalization to the mean free path, 
 $\mfp$, or to the mean crossing time, $\mfp/c$, while
 $I_\nu(\bn)$ is the radiative intensity in the direction of \bn,
%-- (= photon occupancy$/\nu^3$), 
 and $J$ and \bH~ are the first two moments of the intensity.

	Agol and Krolik (1998) 
 computed the first three radiative moment equations of 
 Eq [\ref{eq:Transf}] and, by assuming that multipoles of $I(\bn)$ 
 higher than quadrupole vanish, found an otherwise exact form 
 for the mean intensity perturbation,
 $\delta J \propto 
  	{1 \over \tom} \tilde{\bk} \cdot \delta \bbeta$,  
 in a background with $J = I,~\bH = \emptyset$ 
 (e.g., such that $\delta = \Delta$).
 Tacitly conforming with the premise of isotropy,
 we re-write Eq [23] of Agol \& Krolik (1998) by 
 taking each of the perturbed diagonal components 
 of the radiative field stress tensor, 
 $\delta \underline{K} = 
	(1/4\pi) \delta \, \int d\nu \, d\Omega \, \bn \bn \, I_\nu(\bn)$,
 to be identical and proportional to the perturbed radiative pressure
 $\delta \, \Prad$.  
 Thus
 $\delta K_{ii} = \third \delta J$ and, 
 upon insertion of 
 $-\delta \ln V$ %(c.f., Eqs[\ref{eq:MasterEqs}]) 
in lieu of
%in lieu of\,\footnote{
%Note that this relation is insensitive to the imaginary character of the 
%frequency, e.g., with such interpreted as the growth rate.
%} 
 ${1 \over \tom} \tilde{\bk} \cdot \delta \bbeta$,
 this yields the following relation for quasi-adiabatic 
 perturbations of the radiative component of the fluid
\beq	
	{\delta \Prad \over \Prad}
 	= - {4 \over 3} 
	\left\{ (1 - i\tom) + {i \over 15} \, {\tk^2 \over \tom} \, 
				{{ 5 - 9i\tom}\over{1-i\tom}} \right\}^{-1}
	{ {\delta V} \over V}. 
\label{eq:RadIndex}
\eeq
%-- This equation reflects a general, scale-dependent 
%-- radiative diffusive damping of compressive waves.  
 
	Let us inspect the natural parameterization of this expression.
 When the non-diffusive pressure is negligible, 
 this index represents a truly 
 adiabatic processes for {\it wave} modes only if 
 $\tom, ~\tk^2$ \& $\tk^2/\tom \ll 1$. 
 Individually, both 
 $\tom = \omega (\mfp / c) 
 \simeq |\AM| (\ccH/c) \tau^{-1}_{\rm disk} 
 \simeq 2 \alpha \AM^2 (\ccH/c)^2$,
 and 
 $\tk \simeq k\cH \, \tau^{-1}_{\rm disk} 
 \simeq 2 \sqrt{\alpha} |\AM| (\ccH / c)$, 
 are expected to be small for the
 scales of interest in the magnetorotational instability problem. 
 However, the ratio ~$\tk^2/\tom$~ is not small 
 and this has a non-trivial interpretation:
%-- for semi-isotropic ($\bk_\perp \simeq k_\parallel$) eddies:
\beq
	\left\{ {k^2_\parallel \over \bk^2} \right\}  
 	{\tk^2 \over \tom} 
  	\doteq -i \, 
		\frac	{ \mfp^2 / l^2_{\rm eddy} }
			{ \mfp \, \AOM / c }	
 	\simeq -{i \over |{\rm A}| } 
		  \, \frac	{ (c / \tau_{\rm eddy})}
				{ l_{\rm eddy} } 
	\sim \cO 	\left( 	2i \right)
\label{eq:NoDamp} \eeq
 which identifies with 
 the ratio of eddy diffusion rate to MRI growth rate. 
 \hfill $\alpha_4\varphi\upsilon$

 	To zeroth order, from Eq [\ref{eq:RadIndex}] one thus has 
\beq
 \tilde{\Gamma} \simeq \4ov3 (1 + {\textstyle {i \over 3}} \tk^2/\tom)^{-1}
\label{eq:QuasiAdiab} \eeq
 which indicates that acoustic wave perturbations 
 (i.e., with $\omega$ real)
 on the scale of the MRI are strongly damped if radiative pressure is 
 the predominant source of stress.  
 Indeed, 
 Eq [\ref{eq:QuasiAdiab}] may be interpreted as a form of 
 energy equation characterizing radiative ``heat conduction" out of 
 compressive wave modes (Agol \& Krolik 1998,
 compare this with Eq[51.12] of Mihalas \& Mihalas 1984).

	If there is more parity between the diffusive and non-diffusive 
 sources of stress, we find a generalized quasi-adiabatic index by 
 defining a thermodynamic ``quasi-volume", 
 $\tilde{V} \equiv V^{\teta}$, 
 through Eq [\ref{eq:RadIndex}] as follows
\beq
	{d\Prad \over \Prad} 
	= 	- {4 \over 3} \, \teta  \, {{d V} \over V} 
	\equiv 	- {4 \over 3} \, {{d \tilde{V}} \over \tilde{V}}, 
\label{eq:DiffThermo} \eeq
 while computing 
 the diffusive stress contribution to thermodynamic processes 
 in terms of the logarithmic differential of such quasi-volume:
 $d \ln \tilde{V} = \teta \, d \ln V$.
 (This yields the effective volume filled by the radiative, ``leaky" 
 component of the fluid; the scale-dependent function 
 $\teta$ is nothing but the logarithmic differential ratio 
 between the effective and actual volumes occupied by the gas.)     

 	We consider two cases in turn:
 1- an admixture of photon radiation plus an ideal gas  
 (with $\Pgas \simeq \Prad$) and 
 2- an admixture of neutrinos
 ($\Prad = 7/8 aT^4$) plus gas/radiation/pairs ($\Pgas = 11/12 aT^4$)
 and 
 where the latter species are collisionally coupled at all scales. 
 Furthermore, for neutrino cooled flows we make the simplification that 
 the relatively cool neucleons ($T \ltaprx 20$ MeV) make a negligible 
 contribution to the total pressure (see, e.g., Di Matteo \etal 2002) 

	The standard lore
 (see, e.g.,  Mihalas \& Mihalas 1984, Chandrasekhar 1939)
 computation of the generalized adiabatic exponent, 
 $\Gamma_1$, for an ideal gas plus a radiative component 
 involves setting
 $d Q = d U + d W \doteq \emptyset$ 
 while working out expressions for $d U$ \& $d W$ 
 in terms of two logarithmic differentials, say,
 $d \ln T$ and $d \ln V$.
 Using $d Q \doteq \emptyset$ is clearly artificial but 
 this is an adequate artifact when the main 
 concern is with the non-elastic properties of the fluid
 and not with the amount of heat loss. 
 Thus, we set
 $d Q_{\rm tot} = d Q_{\rm gas} + d \tilde{Q}_{\rm rad} = \emptyset$
 with both $d Q_{\rm gas} = \emptyset$, 
 and $d \tilde{Q}_{\rm rad} = \emptyset$
 (recall that the interactions between the two species are  
 assumed to be entirely elastic).  
 Computation of the specific heat contribution from the gas component 
 is straightforward
 $d Q_{\rm gas}/V =  {\Pgas / (\Gamma_{\rm gas}-1)} d \ln T 
 + \Pgas \, d \ln V$;
 while use of the quasi-volume for the radiative component yields
\beq 
 d \tilde{Q}_{\rm rad}/\tilde{V} 
	= 12 \, \Prad \, d \ln T + 4 \, \Prad \, d\ln \tilde{V} 
	= 12 \, \Prad \, d \ln T + 4 \teta \, \Prad \, d\ln {V}.
\label{eq:QuasiHeatI}
\eeq
	
 	Elimination of $d \ln T$ in favor of $d \ln p$
 is achieved by computing $dp \, (d \ln T,\, d \ln V)$ for the 
 combined gas. 
 Furthermore, utilizing $\Gamma_{\rm gas} = 5/3$
 and the standard definition of 
 $\Gamma_1 \equiv -d_{[\ln V]} \ln p ~(= d_{[\ln \rho]} \ln p)$  
 yields
\beq
	\Gamma_1 = 
	\frac 	{(4\teta\beta+1)(4\beta+1) + 12\beta + 3/2}
		{(1+\beta)(12\beta + 3/2)},	
\label{eq:DmpAdiI} \eeq
 where $\beta \equiv \Prad/\Pgas$.  
 Naturally,
 the $\Re(\Gamma_1)$ agrees with the standard result 
 (Chandrasekhar 1939, Mihalas \& Mihalas 1984).
Moreover, 
it is relatively straightforward to use the 
$\Im(\Gamma_1$) to find the decay rate of acoustic waves partially 
supported by radiative stress (aside from viscosity). 

	For neutrino cooled accretion flows,
 Eq [\ref{eq:QuasiHeatI}] is still valid under the understanding that 
 $\Prad = 7/8 a T^4$ corresponds to the neutrino pressure, while 
 $\Pgas = 11/12 a T^4$ corresponds to photons and pairs and
 $d {Q}_{\rm rad}/{V} 
	= 12 \, \Pgas \, d \ln T + 4 \, \Pgas \, d\ln {V}$.
 With these substitutions, an identical procedure to the one above yields
\beq
	\Gamma_1 =  {4 \over 3}
	\frac 	{(1 + \teta\beta)} {(1+\beta)}.	
\label{eq:DmpAdiII} \eeq

%#########################################################################	
\section{Impact of Radiative Diffusion at MRI Scales}
\label{sec:Impact}
 
	Thus far, 
 we have aimed to maintain the mathematical simplicity afforded 
 by analytical solutions in order to gain physical insight 
 on the phenomenology of radiative heat conduction  
 and on its impact on the MRI.
 
 	We have thus worked out solutions of a simplified version of the
 dispersion relation for non-axisymmetric MRI modes 
 in the limit of purely horizontal fluid displacements.
 Since horizontal displacements maximize the efficiency of free-energy 
 tapping from a differential shear flow,
 this regime generally encompasses the unstable modes of fastest growth.
 The equation of motion, Eq [\ref{eq:k-EOM}],
% which yields the MRI modes of fastest growth, Eq [\ref{eq:qB-om}],
 is applicable to standard heterogeneous fluids with arbitrary 
 generalized index $\Gamma_1$. 
 The ensuing dispersion relation, Eq [\ref{eq:disper}], 
 shows that the MRI is sensitive to a combination of two fluid 
 properties--the ``softness" of the fluid 
 and the strength of the field--through
 a single parameter: $\Lambda$.
 Such parametrization, in turn, bears a direct connection to the
 compressibility of the modes through Eq [\ref{eq:Compress}]. 
 Some physical insight may be gained by re-writing 
 the compressibility parameter,
 $1-\Lambda ~(\propto \Delta \rho / \rho)$ 
 as follows (c.f. Eq [\ref{eq:Compress}]) 
\[ 	1-\Lambda = 
 	\left(1 + {\Gamma_1 \over 2} {\Prg \over \PB } \right)^{-1},
\] 
 while recapitulating that ``2" represents the effective index of a 
 magnetic field to perpendicular compression  (see, e.g., Shu 1992). 
 MRI modes thus involve a large degree of compression when
 the effective sound speed of the material medium
 is much smaller than the \Alf speed.

	On the other hand, 
 in \S \ref{sec:DifAdiaInd} (c.f. Eq [\ref{eq:NoDamp}])
 we have demonstrated that
 when the disk fluid is very hot such that a radiative, diffusive 
 species (photons or neutrinos) constitutes a significant source of 
 stress, radiative heat conduction precludes a formal treatment of 
 the fluid as a single fluid at the scales of relevance to the MRI;
 that is, compression at these scales becomes non-adiabatic. 
 A key conceptual point of this paper is that
 when the diffusive species dominates the stress tensor
 (e.g.,  radiation-pressure--dominated disks),
 the horizontal regime becomes inaccessible because of 
 the brisk loss of pressure support from meridional 
diffusion\footnote{
 Note that the ordering of wavenumber components in the horizontal regime,
 $k_\varphi,  k_r \ll k_\theta$, implies that the smallest length scale
 of the eddies is meridional; i.e., that the eddies are very nearly flat.
}.
 Nevertheless,
 we are looking to read the gross properties of the fastest growth modes 
 from the 2D dispersion relation Eq [\ref{eq:disper}] 
 by requiring parity among diffusive and 
 non-diffusive sources of stress while finding an expression for 
 the {\it generalized quasi-adiabatic index} of the combined fluid,
 Eq [\ref{eq:DmpAdiI} \& \ref{eq:DmpAdiII}].
 It is noteworthy that such parity among sources of stress is to be expected
 in neutrino cooled accretion flows such as those invoked in
 gamma-ray burst progenitor models. 

 	To place a zeroth order 
 (recall $|\tom| \simeq \tk^2 \ll 1,
 ~~\tom / \tk^2 \sim \cO [2i]$)
 upper limit on the quasi-adiabatic index,
% associated with the diffusive species,
%($\tom \simeq \tk^2 \sim \alpha \, \ccH^2/c^2$).
 recall that for wave phase $\propto i(\bk \cdot \bx - \omega t)$,
 the MRI growth rate corresponds to a purely imaginary wave frequency
 (i.e., since the wave phase is stationary at the co-rotation radius):
 $\omega_{_{\rm MRI}} = + i \AOM$.
 Thus, combining Eqs [\ref{eq:NoDamp} \& \ref{eq:QuasiAdiab}] 
 yields
\beq 
 	\tilde{\Gamma} =
		{4 \over 3} \, 
	\frac	{1}{1 + {1 \over 3} \tk^2/\tAOM}
	\longrightarrow
		{4 \over 3} \, 
	\left( 1 + {2 \over 3} 
	\left\{ { \bk^2 \over k^2_\parallel } \right\}  
	\times {\qB^2 \over {\hAOM}}
	\right)^{-1}.
\label{eq:MRIindex} \eeq 	
%-- \label{eq:Rad2MRI} 

 	An accurate estimate of the ratio ${\qB^2 / {\hAOM}}$
 for arbitrary values of $\Lambda (\Gamma_1,\, \Theta)$
 may be found through the iterative use of 
 Eq [\ref{eq:qB-om}] 
 (since $\teta$ and thus $\Gamma_1$ itself depends on ${\qB^2 / {\hAOM}}$),
 but the value of the {\it generalized} index is rather insensitive to 
 small changes in this ratio.  
 For Keplerian rotation and in the weak field 
 (i.e., incompressible) 
 limit, we have 
\[
 {\qB^2 / {\hAOM}}
	\stackrel{\Theta \rightarrow 0}{\longrightarrow} 5/4.
\] 
 On the other hand, the ratio
 $\bk^2 / k^2_\parallel \gg 1$ for 2D motions
 so it should be clear that by setting
 $\bk^2 \geq 2 k^2_\parallel$ 
 we are merely imposing a rough upper limit to the quasi-adiabatic index:
\beq 
 	\tilde{\Gamma} \leq 
		{4 \over 3} \, \teta_{\rm max} 
	\stackrel{\Theta \rightarrow 0}{\longrightarrow}
		{1 \over 2}
\eeq 	
 (again, the result we seek is rather insensitive to the precise value 
 of $\teta_{\rm max}$).  

	Using $\teta_{\rm max} = 3/8$,  
 and $\Prad = \Pgas$, in Eq [\ref{eq:DmpAdiI}],
 the generalized quasi-adiabatic index for an ideal gas plus   
 photon radiation works out to be
\beq
	\Gamma_1 \longrightarrow 1^+.
\label{eq:GenGammaI}
 \eeq
	On the other hand,
 using $\teta_{\rm max} = 3/8$,  
 $\Prad = 7/8 a T^4$ and $\Pgas = 11/12 a T^4$ 
 in Eq [\ref{eq:DmpAdiII}], yields
\beq
	\Gamma_1 = {956 \over 1032} \longrightarrow 1^-
\label{eq:GenGammaII} \eeq
 for a neutrtino cooled accretion flow.
 	In either case, the fluid behavior is close to isothermal
 (in a quasi-adiabatic sense) 
 when the quasi-adiabatic index associated with the diffusive species
 drops below the isothermal value $\tilde{\Gamma} \leq 1$.

%---------------------------------------------------------------------------
\section{Discussion}	

	Our phenomenological discussion is focused on two broad dynamical
 properties of the instability on the presumption that these help clarify
 the physical picture in the non-linear stage. 
 Thus, the growth rate is identified in order of magnitude 
 with the inverse of the correlation time of the turbulence
 and 
 the geometrical regime of fastest growth,
 with the shape of the turbulent eddies.
 We also pay particular attention to the energy deposition process
 at the outer scale since a major portion of this paper is
 centered on the issue of entropy generation through non-adiabatic
 compression when a diffusive species contributes non-trivially to the 
 stress tensor.

	On the basis of the discussion in the previous section, 
 we find that the root of a sluggish growth rate for the MRI 
 (discovered by Blaes \& Socrates 2001 for axisymmetric modes) 
 when radiative pressure {\rm dominates} the non-diffusive component
 is the adiabatic inaccessibility of the horizontal regime   
 associated with fastest growth in standard fluids. 
 For non-axisymmetric modes, 
 this regime becomes intrinsically non-adiabatic 
 because of the brisk rate of heat transport out of 
 anisotropic compressive perturbations.
 Because we begin with a toroidal field configuration 
 where the non-axisymmetric modes are strongly compressive from the onset, 
 the impact of radiative heat conduction is most dramatic and evidenced
 first hand from the dispersion relation.

 	While in an ideal fluid the dominant eddies tend to have  
 $k_\theta$ several times larger than $k_r$ and $k_\varphi$,
 adding a radiative diffusive species will induce 
 conductive losses that move mostly along $\unvc_\theta$. 
 This, in turn, means that the modes with high $k_\theta$,
 i.e., horizontal modes, are preferentially damped.
 The magnitude of $k_\theta$ for the fastest growing modes is then be 
 set by a marginal damping condition.
 Since the length scale for marginal damping is $\approx k_\parallel$
 (c.f. Eq \ref{eq:NoDamp}),
 it follows that 
 MRI eddies should be more nearly isotropic in such an environment, 
 which goes along with the reduction in growth rate.
 For the same reason, 
 energy deposition on to the thermal bath 
 (directly associated with the radiative species, 
 and, secondarily, to the non-radiative species through scattering) 
 occurs on the instability time scale.
 On the other hand, when the source of stress is equally divided among 
 diffusive and non-diffusive particle species,
 the net effect is to soften the effective index of the fluid
 toward isothermality $\Gamma_1 \rightarrow 1$
 (c.f. Eq [\ref{eq:GenGammaI} \& \ref{eq:GenGammaII}])
 along with similar but more modest dynamical effects. 

	These conclusions are broadly supported by the numerical simulations
 of Turner \etal (2001, 2003).  
 Their report of a standard 
 radiation-pressure--dominated $\alpha$-disk, 
 with $\Prad \gtaprx \Pgas \simeq \PB$ shows that 
 the non-linear outcome of the MRI is a porous medium
 with drastic density contrasts.
 Under nearly constant total pressure and temperature, 
 the non-linear regime shows that density enhancements 
 anti-correlate with azimuthal field 
 domains--just as expected from the linear theory--and 
 that turbulent eddies live for about a dynamical time scale
 while matter clumps are destroyed through collisions or by running 
 through localized regions of shear on a similar time scale.
 Notably, the non-linear density contrasts may be quite large
 $<\rho_{\rm max}/\rho_{\rm min}> \gtaprx~ 10$ 
 when $\Prad \gg \Pgas$.   

 	At a very fundamental level,
 clump formation is intimately connected to the effects of 
 radiative heat conduction out of compressive perturbations.
 This can be understood by computing the polarization properties 
 of the fastest growing modes in the horizontal regime:
 $\xi_r = - \sqrt{\Lambda} ~\xi_\varphi~\&~|\xi_\theta| \ll |\xi_r|$,
 and reading from Eq [\ref{eq:Compress}] that 
 at the linear stage of the instability there exists 
 a converging flow toward the Lagrangian displacement node of the modes
 (see Fig. 2 of Foglizzo \& Tagger 1995). 
 In a fluid with entirely elastic (adiabatic) properties, 
 the pressure perturbation associated with such compression will 
 act as a restoring force to de-compress the fluid in the non-linear stage.   
 On the other hand, 
 when the fluid has radiative heat conduction properties 
 on the scale of the density perturbations, 
 no such restoring force persists on time scales longer than 
 the inverse of the growth rate 
 (which, following Eq [\ref{eq:NoDamp}],
 is of the same order as the diffusion time).
 Material clumps thus formed will maintain their integrity and
 survive for longer times.
 When the diffusive and non-diffusive pressures are comparable,
 these arguments still hold but only in part.
 In this case,
 compressive perturbations in the non-linear stage will decompress
 only in proportion to the remaining pressure support from 
 the non-diffusive species.

%	Energy injection at the outer scale by the MRI causes
% a turbulent cascade of energy in fluctuating magnetic and velocity fields.
% Since the typical speed of MRI motions is the \Alf speed,
% the cascade in a standard fluid proceeds mostly along the two channels
% that most nearly match this speed:
% an incompressible Alfvenic and a compressible pseudo-Alfvenic channel.
% For a radiative fluid, however, diffusive heat transport effectively
% eliminates the latter channel from the cascade since compressive waves
%-- solely supported by radiative pressure on the scale of the MRI 
% at the outer scale are strongly suppressed
% (c.f. Eq [\ref{eq:QuasiAdiab}] with $\hat{\bk} \simeq \unvc_{\hat{\bB}}$,
% yields 
% $\Re(\Gamma_1) \simeq -\third \Im(\Gamma_1)$)

	Blaes \& Socrates (2001) predicted qualitatively similar results
 for axisymmetric modes.  Our analytical solutions agree with their result, 
 $\hAOM \sim c_g/\Valf$, 
 in the limit $k_r = c_r \rightarrow 0$ and $c_g \rightarrow 0$. 
 This paper complements these findings
 and the non-linear numerical explorations of Turner \etal (2001, 2003)
 by presenting a simplified approximate dispersion relation 
 for non-axisymmetric modes 
 while interpreting the effects of radiative heat conduction 
 in terms of an ``ultra-soft" quasi-adiabatic index 
 (i.e., $\tilde{\Gamma} < 1$)
 for the radiative contribution to the stress tensor.
 The quasi-adiabatic index and all derived thermodynamic quantities
 (including ``quasi-specific" quantities that ensue from division by
 the quasi-volume, \S \ref{sec:DifAdiaInd})
 are mere mathematical artifacts to manipulate
 thermodynamical relations in the usual manner. 
% Foreseen caveats are discussed in \S \ref{sec:DifAdiaInd}. 
 Furthermore, we have demonstrated that when there is parity among 
 the sources of stress, the combined gas behaves isothermally.
 Our results are generally applicable to optically thick, 
 neutrino-cooled disks 
 since the pressure contributions from photons, pairs and neutrinos, 
 all have the same temperature dependence 
 whereas only the neutrino component has radiative heat conduction 
 properties on the time and length scales of the instability.
 
%- Since this results in a convergent flow at the displacement nodes,
%- one may expect these sites to be ripe ground for Fermi %--second order 
%- acceleration %-- of both photons and baryons 
%-- (R. Blandford, Priv. Comm.).  	

%########################################################################
\section*{Acknowledgments}
 	R.A.G. would like to acknowledge discussions with Omer Blaes 
 and the hospitality of 
 Caltech's Theoretical Astrophysics and Relativity Group. 

%########################################################################
 
\end{document}